\documentstyle[multicol,aps,epsfig]{revtex}
\begin{document}
\draft
\title{Current self-oscillations, spikes and crossover between
charge monopole and dipole waves in semiconductor superlattices}
\author{David S\'anchez$^1$, Miguel Moscoso$^{2,3}$,
        Luis L. Bonilla$^{2,3}$, Gloria Platero$^1$ and
        Ram\'on Aguado$^1$}
\address{$^1$Instituto de Ciencia de Materiales (CSIC),
Cantoblanco, 28049 Madrid, Spain\\
$^2$Escuela Polit{\'e}cnica Superior,
Universidad Carlos III de Madrid,
Avenida de la Universidad 20, \\ 28911 Legan{\'e}s, Spain. \\
$^3$Also: Unidad Asociada al Instituto de Ciencia de Materiales
(CSIC)}
\date{\today}
\maketitle
\begin{abstract}
Self-sustained current oscillations in weakly-coupled 
superlattices are studied by means of a self-consistent 
microscopic model of sequential tunneling
including boundary conditions naturally. Well-to-well
hopping and recycling of charge monopole domain walls
produce current spikes --high frequency modulation--
superimposed on the oscillation.
For highly doped injecting contacts, the self-oscillations
are due to dynamics of monopoles. As the
contact doping decreases, a lower-frequency oscillatory mode due to
recycling and motion of charge dipoles is predicted. For low
contact doping, this mode dominates and monopole oscillations
disappear. At intermediate doping, both oscillation modes coexist
as stable solutions and hysteresis between them is possible.

\end{abstract}

\begin{multicols}{2}
\narrowtext

Solid state electronic devices presenting negative differential 
conductance, such as resonant tunneling diodes,
Gunn diodes or Josephson junctions~\cite{book},
are nonlinear dynamical systems with many degrees of freedom.
They display typical nonlinear phenomena such as
multistability, oscillations, pattern formation or bifurcation 
to chaos. In particular, vertical transport in weakly
coupled semiconductor doped superlattices (SLs) has been
shown to exhibit electric field domain 
formation~\cite{cho87,gra91,bon95}, multistability~\cite{kas94},
self-sustained current oscillations~\cite{mer95,kas97,bon97}, and
driven and undriven chaos~\cite{bul95,zha96,luo98}. Stationary
electric  field domains appear in voltage biased SLs if the doping
is large enough~\cite{bon95}. When the carrier density is below a
critical value, self-sustained oscillations of the current may 
appear. They are due to the dynamics of the domain wall (which 
is a charge monopole accumulation layer or, briefly, a {\em
monopole}) separating the electric field domains. This domain wall
moves through the structure and is periodically recycled. The
frequencies of the corresponding oscillation depend on the applied
bias and range from the  kHz to the GHz regime. Self-oscillations
persist even at room temperature, which makes these devices
promising candidates for microwave generation~\cite{kas97}.
Theoretical and experimental work on these systems have gone hand
in hand. Thus the paramount role of monopole dynamics has been
demonstrated by theory and experiments. Monopole motion and
recycling can be experimentally shown by counting the spikes --high
frequency modulation--  superimposed on one period of the current
self-oscillations: current spikes correspond to well-to-well
hopping of a domain wall through the SL. In typical experiments
the number of spikes per oscillation period is clearly less than
the number of SL wells~\cite{kas97,kan97}. It is known that
monopoles are nucleated well inside the SL~\cite{kas97,bon97}
so that the number of spikes tells over which part of the
SL they move. Other possible waves, such as the charge
dipole waves appearing in the well-known Gunn effect, are nucleated
at the emitter contact~\cite{hig92}. Had they been mediating the
self-oscillation, the number of current spikes would be
comparable that of SL wells. 

In this letter we study the non-linear dynamics of SLs by 
numerically simulating the model proposed in Ref.\
\onlinecite{agu97}. Our simulations show self-sustained 
oscillations of the current and current spikes 
reflecting the motion of the domain wall
as observed experimentally. Furthermore, when contact
doping is diminished, we predict a crossover from
monopole to dipole self-oscillations resembling
those in the Gunn effect~\cite{hig92}.
Indeed, our results show for first time that 
there is an intermediate range of contact
doping and a certain interval of external dc voltage for 
which monopole and dipole self-oscillations with different 
frequencies are both stable. Hysteretic phenomena
then exist. 

\noindent {\em 1. Model and superlattice sample}. Our
self-consistent microscopic model of sequential tunneling 
includes a detailed electrostatic description of the contact 
regions and SL~\cite{agu97}. It consists of a system of $3N+8$
equations for the Fermi energies and potential drops at the $N$
wells, the potential drops at the barriers and 
at the emitter and contact layers, width thereof, charge at the
emitter and total current density. These equations comprise
Amp\`ere current density balance and Poisson equations, 
conservation of the global charge, and the overall voltage
bias condition. Dynamics enters the model through Amp\`ere's
law for the total current density $J=J(t)$,
\begin{eqnarray}
J =  J_{i-1,i} + {\epsilon\over d}\, {dV_{i}\over dt}\,,
\label{1}
\end{eqnarray}
which is equivalent to a local charge continuity
equation~\cite{agu97}.
Here $J_{i-1,i}$ is the tunneling current density
through the $i$th barrier of thickness $d$, evaluated
by using the Transfer Hamiltonian approach~\cite{agu97}.
The last term in (\ref{1}) is the displacement current
at the $i$th barrier where the potential drop is $V_{i}$ and
$\epsilon$ is the static permittivity.

Our numerical simulations (of the $3N+8$ coupled equations)
have been performed for a 13.3~nm GaAs/2.7~nm AlAs 
SL at zero temperature consisting of 50 wells and 51 barriers,
as described in \cite{kan97}. Doping in the wells and in the 
contacts are $N_{w} = 2\times 10^{10}$~cm$^{-2}$ and $N_{c}= 2
\times 10^{16}$~cm$^{-3}$ respectively. Notice that the typical
experimental value is $N_c = 2\times 10^{18}$~cm$^{-3}$
\cite{kas97,kan97}. For this value, we find current
self-oscillations due to monopole dynamics with very small
superimposed current spikes. Since the origin of such spikes is the
same as for smaller $N_c$ (for which spikes are larger and
bistability of oscillations is possible), we choose not to present
data comparable to experiments in this paper (see Ref.\
\onlinecite{kan97} for the relevant experimental data). 

\begin{figure}[!htp]
\centerline{
\epsfig{file=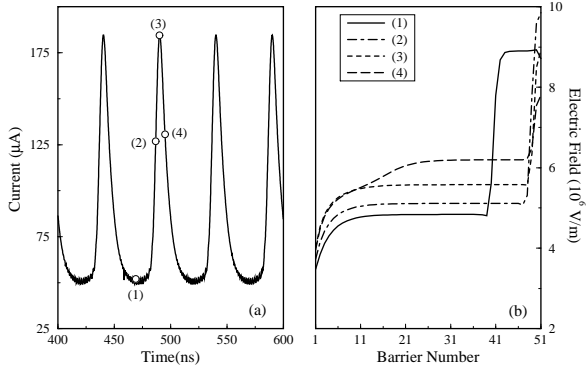,angle=270,width=0.45\textwidth}
}
\caption{(a) Self-sustained oscillations of the total current
through the SL due to monopole recycling and motion. Bias
is $5.5 V$ and emitter doping, $N_{c}=2\times
10^{16}$~cm$^{-3}$. (b) Electric field profiles at the times
marked in (a) during one period of the current oscillation.}
\label{monosc}
\end{figure}

\noindent {\em 2. Monopole-mediated self-oscillations of the
current}.
Fig.~\ref{monosc}(a) depicts the current as a function of time
for a dc bias voltage of 5.5~V on the second plateau of the SL
$I-V$ characteristic curve. $J(t)$ oscillates periodically at
20~MHz. Between each two peaks of $J(t)$, we observe 18
additional spikes. The electric field profile is plotted in
Fig.~\ref{monosc}(b) at the four different times of one
oscillation period marked in Fig.~\ref{monosc}(a). There are two
domains of almost constant electric field separated by a moving
domain wall of (monopole) charge accumulation (which is extended
over a few  wells). Monopole recycling and motion occur
on a limited region of the SL (between the 30th and the 50th
well) and accompany the current oscillation~\cite{kas97,bon97}.
Well-to-well hopping of the domain wall is reflected by
the current spikes until it reaches the 46th well which
is close to the collector. Then the strong influence of the contact
causes that no additional spikes appear.
Instead the current rises sharply triggering the formation of a new
monopole closer to the emitter contact but well inside the SL; see
Figs.~\ref{monosc}(a)  and (b). The number of wells traversed by
the domain wall (almost) coincides with the
number of spikes per oscillation period, {\em a feature not
found in previous models}. Fig.~\ref{monosc}(b)
shows the recycling of a monopole: between times (1) and (3)
there is a single monopole propagating towards the collector; at
(4) a new monopole is generated at the middle of the structure and
the old one collapses at the collector. It is interesting to realize
that the region near the emitter does not have a constant electric
field profile due to the large doping there (its Fermi level is
well above the first resonant level of the first well). This
produces a large  accumulation layer.

\noindent {\em 3. Current spikes}.
What is remarkable in Fig.~\ref{monosc}(a) (as
compared to previous studies) are the spikes superimposed near
the minima of the current oscillations.
Such spikes have been observed experimentally and attributed
to well-to-well hopping of the domain wall~\cite{kan97,kas96}. 
They are a cornerstone to
interpret the experimental results and in fact support the
theoretical picture of monopole recycling in part (about $40\%$)
of the SL during self-oscillations. The identification
between number of spikes and of wells traversed by the
monopole rests on voltage turn-on measurements supported
by numerical simulations of simple models during early stages
of stationary domain formation~\cite{kas96}. These models 
do not predict spikes superimposed on current self-oscillations 
due to monopole motion~\cite{bon95,kas97,wac97}. To predict 
large spikes, a time delay in the tunneling current~\cite{kan97} or
random doping in the wells~\cite{pre96} have to be added. Unlike
these models, ours reproduces and explains spikes naturally
thereby supporting their use to interpret experimental results. 

\begin{figure}[!htp]
\centerline{
\epsfig{file=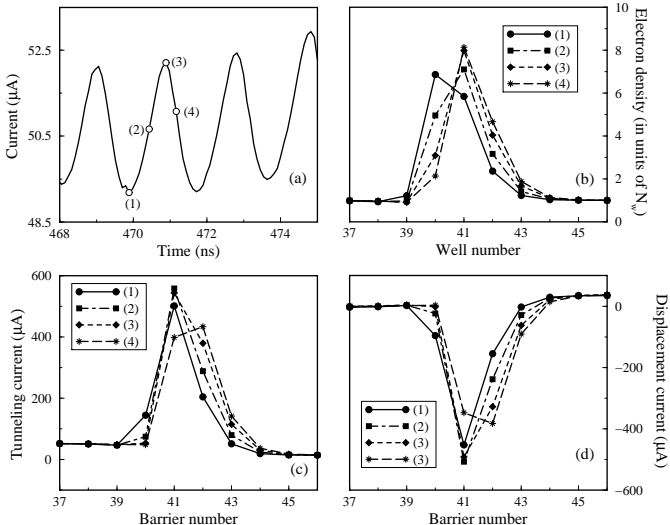,angle=270,width=0.45\textwidth}
}
\caption{(a) Zoom of Fig.~\ref{monosc} showing the spikes of the
current. (b) Electron density profiles (in units of the
doping at the wells), (c) tunneling current,
and (d) displacement current within the monopole
at the times marked in (a).}
\label{spikes}
\end{figure}

Fig.~\ref{spikes}(a) depicts a zoom of the spikes in 
Fig.~\ref{monosc}(a). They have a frequency of about 
500~MHz and an amplitude of 2.5~$\mu$A. Fig.~\ref{spikes}(b)
shows the charge density profile at four different times of a
current spike marked in Fig.~\ref{spikes}(a). Notice that the
electron density in Fig.~\ref{spikes}(b) is larger than the well
doping at only three wells (40, 41 and 42) during the times
recorded in Fig.~\ref{spikes}(a). The maximum of electron
density moves from well 40 to well 41 during this time interval
so that: (i) tunneling through the 41st barrier (between wells 40
and 41) dominates when the total current density is increasing,
whereas (ii) tunneling through barriers 41 and 42 is important
when $J(t)$ decreases. The contributions of tunneling and
displacement currents to $J(t)$ in Eq.\ (\ref{1}) are depicted
in Figures \ref{spikes}(c) and (d).

More generally, the spikes reflect the two-stage hopping
motion --fast time scale-- of the domain wall: at time (1) 
(minimum of the current), the charge accumulates mainly at the 
{\it i-th} well. As time elapses, electrons tunnel from
this well to the next one, the {\it (i+1)-st}, where most of the
charge is located at time (3) (maximum of the current). This 
corresponds to a hop of the monopole. As the monopole moves, it 
leaves a lower potential drop on its wake.
The reason is that the electrostatic field at the {\it 
(i+1)-st} well and barrier become abruptly flat between times
(1) and (3), as they pass from the high to the low field domain.
This means that a negative displacement current has its 
peak at the {\it (i+1)-st} barrier, near the wells where most 
of the charge is. Between times (1) and (3),
the tunneling current is maximal where
the displacement current is minimal and the total current
increases. After that, some charge flows to the next well
[time (4)] but both, tunneling and displacement currents, are
smaller than previously. This occurs because 
the potential drop at barrier {\it (i+2)} (in the high
field domain) is larger than at barrier {\it (i+1)}.
Then there is a smaller overlap between the resonant levels
of nearby wells --the tunneling current decreases -- and 
the displacement current and, eventually, $J(t)$ decreases.
This stage lasts until well {\it i} is drained, and most of the
charge is concentrated at wells {\it (i+1)} (the local maximum
of charge) and {\it (i+2)} (slightly smaller charge). Then the
next current spike starts.

\begin{figure}[!htp]
\centerline{
\epsfig{file=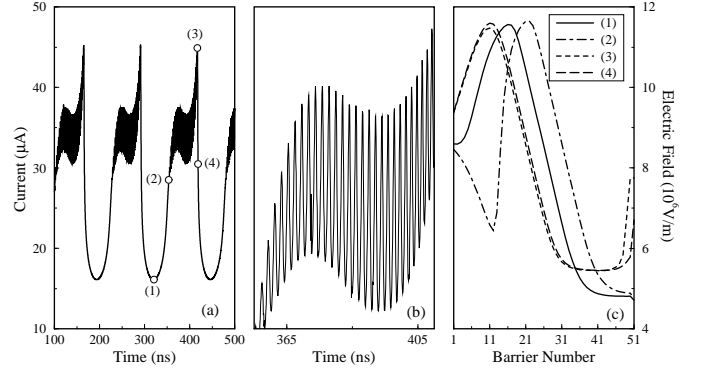,angle=270,width=0.45\textwidth}
}
\caption{(a) Dipole-mediated self-oscillations of the current at
$5.5 V$ for $N_{c}=2\times 10^{16}$cm$^{-3}$.
(b) Detail of the current spikes. (c) Electric field profiles at
the times marked in (a).}
\label{diposc}
\end{figure}

\noindent {\em 4. Dipole self-oscillations of the current}.
An advantage of our present model over other discrete
ones~\cite{bon95,wac97,pre94} is our microscopic modeling of
boundary conditions at the contact regions. Thus we can study what
happens when contact doping is changed. The result is that there
appear dipole-mediated self-oscillations as the emitter doping is
lowered below a certain value.
There is a range of voltages for which dipole and monopole oscillations 
coexist as
stable solutions. This range changes for different plateaus.
When the emitter doping is
further lowered, only the dipole self-oscillations remain.
Fig.\ref{diposc} presents data in the crossover range 
(below $N_c = 4.1\times 10^{16}$cm$^{-3}$
and above $N_c = 1.7\times 10^{16}$cm$^{-3}$ for the second plateau),
for the same sample, doping and bias as in Figs.\
\ref{monosc} and \ref{spikes}. Except for the presence of spikes
of the current, dipole recycling and motion in SLs are similar to
those observed in models of the Gunn effect in bulk
GaAs~\cite{hig92}. These self-oscillations have not been observed
so far in experiments due to the high values of the contact doping
adopted in all the  present experimental settings.
Notice that current spikes appear differently than in the monopole
case, Fig.~\ref{monosc}(a). The main difference is that now there
are many more current spikes, 36, for the dipoles recycle at
the emitter and traverse the whole SL. See Figs.~\ref{diposc}(b)
and (c). Charge  transfer and balance between tunneling and
displacement current during a spike are similar to those occurring
in monopole oscillations. For a simpler model~\cite{bon95,bon97}
the velocity of a charge accumulation layer (belonging to a monopole
or a dipole) has been shown to approximately obey an equal area
rule. Then monopole and dipole velocities
are similar but a monopole traverses a smaller part of the
SL than a dipole does. Therefore dipole oscillations have
a lower frequency than monopole ones. Our results agree
with this: the frequency of the dipole oscillations discussed
above is about 8~MHz, 40\% the frequency of monopole oscillations.

Dipole self-oscillations have also been predicted to occur in 
weakly-coupled SLs as the result of assuming a linear current --
field relation at the injecting contact on a simpler
model~\cite{wac97,bon98}.
Since such {\em ad hoc} boundary condition has no 
clear relation to contact doping, no crossover between 
different oscillation types could appear in that work. 

\begin{figure}[!htp]
\centerline{
\epsfig{file=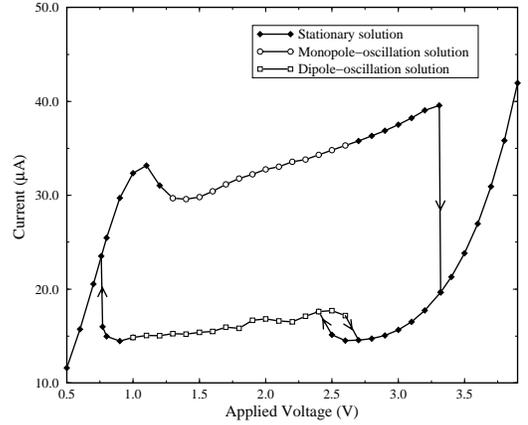,angle=270,width=0.45\textwidth}
}
\caption{I-V characteristics at the first plateau,
for both sweep directions
showing bistability between self-oscillations
mediated by monopole and by dipole dynamics.
Notice the hysteresis cycle.    }
\label{biest}
\end{figure}

\noindent {\em 5. Multistability}. Monopole and dipole waves
coexist in both the first and the second plateaus.
The time-averaged current as a function of dc voltage in
the first plateau (whose crossover range is below
$N_c = 2.1\times 10^{16}$cm$^{-3}$
and above $N_c = 1.5\times 10^{16}$cm$^{-3}$)
has been plotted in Fig.~\ref{biest}. Notice
that the average current of dipole oscillations is lower than
that of monopole oscillations. Previous studies for
Gunn oscillations~\cite{hig92} found
that large dipole waves appear only for small current values,
whereas monopole recycling requires current values near the maximum of the
current-field characteristic curve. Let us
start at a bias of 0.5~V (for which the  stationary state is
stable) and adiabatically increase the voltage. The result is that
we go smoothly from the stationary state to the fast monopole
self-oscillation at about $1.3$~V. This branch of oscillatory states
eventually disappears at about 2.6~V. If we now
adiabatically  lower the bias, we reach a slow dipole
self-oscillation at about 2.4~V. There is a small hysteresis loop
between dipole oscillations and the stationary state between
2.4~V and 2.6~V: the former may start as a subcritical
Hopf bifurcation.
About 0.8~V the dipole oscillation disappears and
we are back at the stable stationary state. We therefore find the
hysteresis loops marked by arrows in Fig.~\ref{biest}.

In conclusion, we have dealt with self-sustained oscillations
of the current in SLs whose main mechanism is sequential
tunneling. Depending on contact doping, these oscillations may
be due to recycling and motion of two different charge density
waves: monopoles and dipoles. Experimentally, only the monopole
oscillations have been observed, for the contacts doping is usually
set to values which are too high. The dipole-like oscillations
could be  observed constructing samples with lower
doping at the contacts. In fact, as the doping of the contacts is
reduced, we predict current oscillations due to dipole charge
waves. The crossover between both types of self-oscillations occurs
at intermediate emitter doping values for which stable monopole and
dipole oscillations coexist. Then the diagram of average current
versus  voltage is multivaluated, presenting hysteresis cycles and
multistability between monopole and dipole oscillations (and
between oscillatory and stationary states). The time-resolved
current in the oscillatory modes presents a number of sharp spikes.
They occur because well-to-well hopping of charge accumulation
layers occurs in two stages: during the stage where the current
rises, charge is mainly transferred through a single  barrier. The
charge is transferred through two adjacent barriers at the stage in
which the current decreases. All these properties
form the basis for possible applications of SLs working as
multifrequency oscillators in a wide range of frequencies. 
Quantitative description of such multifrequency oscillators 
requires calculation of typical output power characteristics and
noise levels. This is the purpose of a future work.\\

{\bf Acknowledgments}. We thank Rosa~L\'opez for helpful
discussions. This work has been supported by the DGES (Spain)
grants PB97-0088, PB95-1203 and PB96-0875, by the European Union
TMR contracts ERB FMBX-CT97-0157 and FMRX-CT98-0180 and by the Community
of Madrid, project 07N/0026/1998.

\end{multicols}
\end{document}